%% file: main.tex
\documentclass[runningheads]{llncs}
\usepackage[T1]{fontenc}
\usepackage{cite}
\usepackage{algorithmic}
\usepackage{graphicx}
\usepackage{textcomp}
\usepackage{xcolor}
\usepackage{paralist}
\usepackage{balance}
\usepackage{booktabs}

\usepackage{amsthm}
\usepackage{amsmath,amssymb,amsfonts}
\usepackage{diagbox}

\PassOptionsToPackage{hyphens}{url}
\usepackage{hyperref}

\DeclareMathOperator*{\argmax}{arg\,max}    
\newtheorem{The}{Theorem}

\newtheorem{Pro}[The]{Proposition}

\newcommand{\sln}{{\it zkRansomware}}
\newcommand{\slntitle}{zkRansomware}

\newcommand{\para}[1]{\smallskip\noindent\textbf{#1}~}
\newcommand{\parait}[1]{\smallskip\noindent\textit{#1}~}

\begin{document}

\title{zkRansomware: Proof-of-Data Recoverability and Multi-round Game Theoretic Modeling of Ransomware Decisions}

\author{Xinyu Hou\inst{1} \and
Yang Lu\inst{2}\orcidID{0009-0006-0981-7211} \and
Rabimba Karanjai\inst{2}\orcidID{0000-0002-6705-6506} \and
Lei Xu\inst{3}\orcidID{0000-0002-7662-2119} \and
Weidong Shi\inst{2}\orcidID{0000-0002-1994-4218}}
\authorrunning{X. Hou et al.}
%
\institute{University of Science and Technology of China, Anhui, 230052, China \\
\email{houxinyu123@mail.ustc.edu.cn}
\and
University of Houston, Houston, TX 77204, USA \\
\email{ylu17@central.uh.edu} \\
\email{rkaranja@cougarnet.uh.edu} \\
\email{larryshi@ymail.com}
\and
Kent State University, Kent, OH 44240, USA\\
\email{xuleimath@gmail.com}}

\titlerunning{zkRansomware}
\maketitle

\begin{abstract}
    Ransomware is still one of the most serious cybersecurity threats. Victims often pay but fail to regain access to their data, while also facing the danger of losing data privacy. These uncertainties heavily shape the attacker-victim dynamics in decision-making. In this paper, we introduce and analyze \sln{}. This new ransomware model integrates zero-knowledge proofs to enable verifiable data recovery and uses smart contracts to enforce multi-round payments while mitigating the risk of data disclosure and privacy loss. We show that \sln{} is technically feasible using existing cryptographic and blockchain tools and, perhaps counterintuitively, can align incentives between the attacker and the victim. Finally, we develop a theoretical decision-making framework for \sln{} that distinguishes it from known ransomware decision models and discusses its implications for ransomware risk analysis and response decision support. 
\end{abstract}

\input{Introduction}
\input{design}
\input{context}
\input{simulations}
\input{mitigation}

\input{conclusion}

\balance
\bibliographystyle{IEEEtran}
\bibliography{ref}

\input{appendix}

\end{document}

%% file: Introduction.tex
\section{Introduction}

Ransomware has become one of the most popular types of malicious software~\cite{10.1145/3514229,10.1145/3691340,o2018evolution,gazet2010comparative}.  
This malicious software encrypts the victim's critical data, making it impossible to access the system and retrieve the data. It then demands a ransom from the victim through a cryptocurrency system to restore system functionality and data files. 
In the past decade, ransomware has experienced significant development and even held the crown as the fastest-growing cybersecurity threat. 
The ransomware launch rate has increased exponentially from a new attack every 40 seconds in 2016 to every 16 seconds in 2019~\cite{morgan2019ransomware,kaspersky2016security}. 
By 2031, the estimated cost associated with ransomware attacks will reach \$275 billion~\cite{ransomcost}. According to a report by Sophos, in 2023, approximately 73\% of companies worldwide paid ransoms to recover data after a ransomware attack. 
Paying the ransom does not guarantee data recovery. According to Sophos Cybersecurity report, during ransomware attacks, 
53\% of the ransomware victims who paid the ransom did not get their data back~\cite{6reasonsnottopay}. 

Traditional ransomware encrypts information on a victim’s computer to demand a ransom payment for the decryption key, which is modeled as Ransomware 1.0.
The attacker only demands a ransom and decides whether to return the decryption key to the victim or not. The victim only needs to worry about whether the data can be recovered.  
Ransomware 1.5 introduced the data threat ransom. Building on Ransomware 1.0, attackers can publicly release the victim's data, causing even more damage to the victim who refuses to pay the ransom, thus increasing their ``willingness to pay''. 
In Ransomware 2.0~\cite{li2021game,li2020ransomware}, attackers can sell the victim's data for extra profit, making these ransomware attacks more lucrative and leaving the victim increasingly vulnerable. 

With technology advance, ransomware attacks will certainly evolve ~\cite{10.1145/3691340}, which motivates the investigation of the trends of ransomware attacks and the applicability of various emerging technologies in the context of ransomware attack and defense. The goal is threefold:
\begin{inparaenum}[\bfseries (i)]
    \item to explore the emerging and possible evolution path of ransomware attacks;
    \item to avoid future surprises; and 
    \item to provide analysis model and game theory based decision support tools to assist the victims. 
\end{inparaenum}

We observe that a new type of ransomware attack may emerge leveraging already available technology components. This new breed of ransomware attack is worth close study because it alters the well-established decision modeling developed for understanding decision making process in ransomware attacks. We call this new type of ransomware attack, \sln{}, built on top of zero-knowledge protocols (ZKP~\cite{goldreich1994definitions}) and smart contracts~\cite{wang2018overview}, which allows a victim to verify whether the data can be recovered before making a payment and to split the payment into multiple rounds with smart contract based guarantee to reduce the risk of data privacy loss.
Through game theoretical decision modeling, we show that this new attack scenario offers a significant incentive to the ransomware attacker for adoption. 
On the other hand, it also provides certain benefits to the victims, such as improved assurance of data recovery and enhanced risk management options to data privacy loss. 
This indicates that \sln{} points to a new decision-making equilibrium between the ransomware attacker and the victim, beyond the current ransomware decision models. 

In summary, the main contributions of our work are as follows: 
\begin{inparaenum}[\bfseries (i)]
    \item We describe the details of a new type of ransomware attacks, \sln{}, which is built on top of the existing and mature technology components, including verifiable encryption, fair data exchange, and smart contracts. The potential integration of these components by the attackers can enable a new ransomware ecosystem that demands understanding in research. 
    \item We present a new game-theoretical model for analyzing the decisions under this new ransomware framework. The decision model contributes to understanding emerging threats, evaluates the impacts of adopting new technology components by ransomware attacks, 
    and keeps the community informed. 
    \item We provide analysis and simulation experiments to show different trade-offs under \sln{} and its decision model. 
\end{inparaenum} 

It is important to point out that the study is motivated to understand and model ransomware decision process in the face of emerging threats. 

%% file: design.tex
\section{zkRansomware}\label{sec-ransomware-framework}
This section presents \sln{}, an emerging ransomware threat that utilizes zero-knowledge succinct non-interactive arguments of knowledge (zk-SNARK) ~\cite{cryptoeprint:2017/540,cryptoeprint:2016/260,cryptoeprint:2018/187} and smart contract technologies atop the existing ransomware attacks.

\para{Key enabling technologies: fair data exchange and verifiable encryption}
Fair Data Exchange (FDE) protocols ensure secure and equitable data transactions between untrusted parties~\cite{atomicfairdataexchange}. In these protocols, a client locks the payment in a smart contract, and the server receives funds only upon releasing a valid decryption key that matches a committed data hash ~\cite{atomicfairdataexchange}.
Recent FDE designs use zero-knowledge proofs and verifiable encryption over committed keys (VECK), enabling constant-size commitments via schemes such as KZG~\cite{10.1007/978-3-642-17373-8_11}. These techniques also support data retrievability and replication proof ~\cite{10.1145/1655008.1655015,10.1007/978-3-030-17656-3_12}.

Verifiable encryption~\cite{10.1007/978-3-540-45146-4_8} targets the problem of proving the properties of encrypted data. 
Verifiable encryption by itself does not enable FDE. However, it may provide a foundation when customized to work with other technology components to support fair exchange of encrypted data.

\para{Security assumptions}
To analyze \sln{}, the security assumptions are as follows:
\begin{inparaenum}[\bfseries (i)]
    \item \textit{Rationality of the participants in the ransomware decision game}
        Both the attacker and the victim are rational players. The attacker is financially motivated, seeking to maximize its profits through ransomware attacks. This means that the attacker when facing action options, will pick the action that can lead to higher chance of receiving more ransom. 
        The attacker may sell the victim’s data if doing so is profitable. Under certain circumstances, the victim is willing to pay the ransom to regain access to the data. Furthermore, we assume that the victim cares about data confidentiality. If the victim has options, he/she would prefer the option to minimize the risk of losing data privacy or delay data leakage if possible. The security model excludes the case where the victim intentionally breaks data confidentiality and leaks his/her own data. 
    \item \textit{Blockchain as a trusted 3rd party.}
        The blockchain used for smart contracts deployment and enforcement is trusted. Fair exchange is impossible without a trusted third party~\cite{Pagnia1999OnTI}. In this case, blockchain acts as a trusted third party. 
    \item \textit{FDE security.} It is assumed that the cryptographic schemes in FDE protocols, such as verifiable encryption and cryptographic commitments (e.g., KZG polynomial commitment), are secure. The attacker applies verifiable encryption over committed keys (VECK) to encrypt the victim’s data.  
    \item \textit{Availability of data required for proof verification.} 
        After a successful ransomware attack, the attacker will remove the original data and the decryption key from the victim’s computer. However, the attacker will leave necessary data for the victim to verify in a FDE protocol. The attacker is very well motivated to do so because it increases the chance to receive ransom.
\end{inparaenum}

\begin{figure*}
\centering
        \includegraphics[height=4.4cm,width=6.8cm]{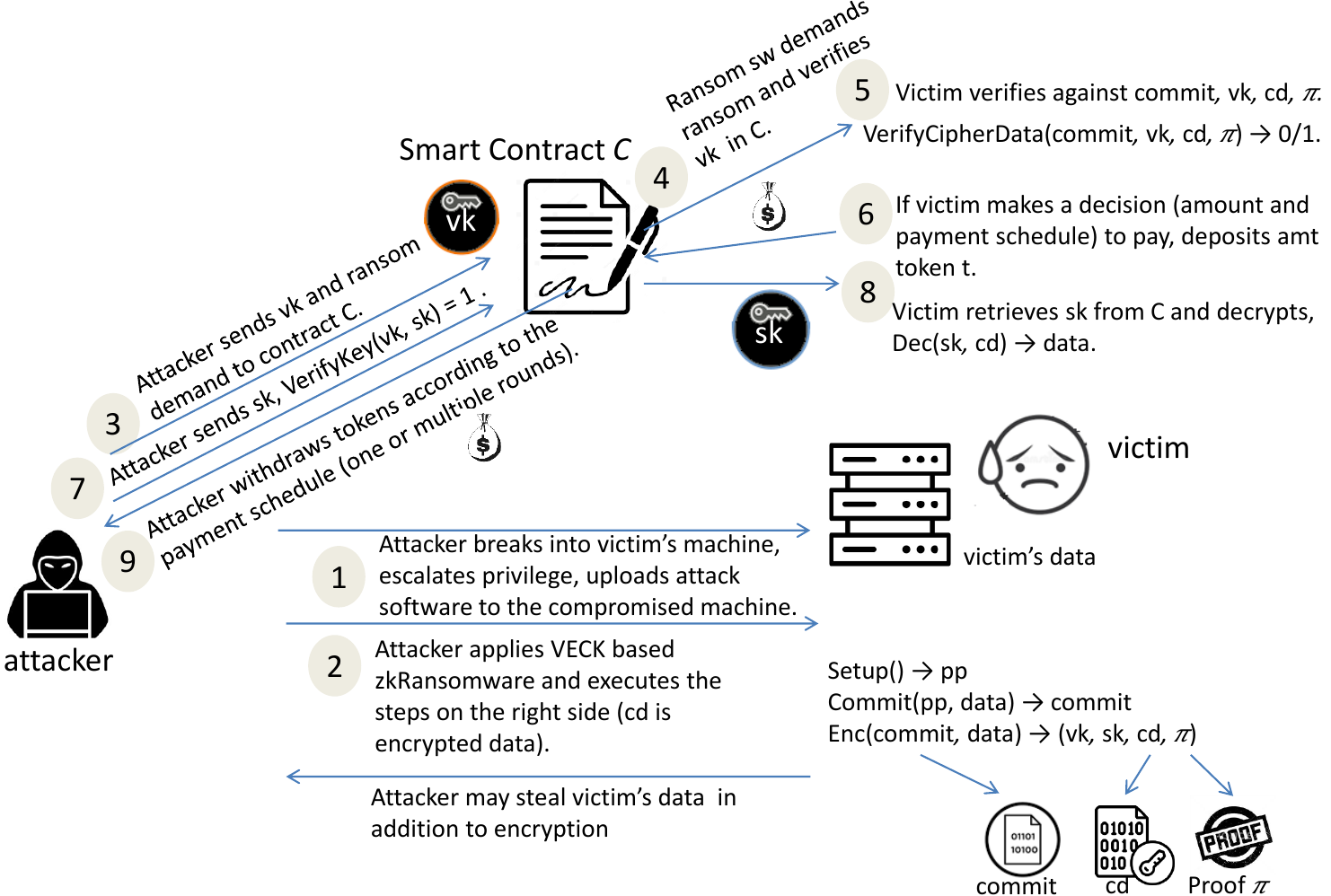}
        \vspace{-0.15in}
        \caption{A high-level diagram of \sln{}.}
        \label{fig:protocol}
        \vspace{-0.2in}
\end{figure*}

\para{Description of \slntitle{}}
There are three entities involved:
\begin{inparaenum}[\bfseries (i)]
    \item The attacker $A$ who launches ransomware attacks;
    \item The victim $V$ who is the target of a ransomware attack; and
    \item The blockchain $B$ that hosts smart contract $C$ that is used to support attack/payment related operations.
\end{inparaenum}


\parait{Single round payment}
\figurename~\ref{fig:protocol} demonstrates the major steps of a ransomware attack with a single round payment.
Specifically, the entities interact with each other through the following steps:
\begin{inparaenum}[]
    \item \textbf{Step 1.} Attacker $A$ breaks into victim $V$'s computer to gain privileged access and uploads the attacking code to the victim’s computer. 
    \item \textbf{Step 2.} Attacker $A$ applies a VECK-based scheme to encrypt selected data on the victim’s machine. This involves a sequence of actions. \texttt{Setup()} creates security parameters required by the VECK scheme. Then, the attacker computes the commitment value, $commit$, based on the input data. The attacker runs \texttt{Enc()} to encrypt the data, and the results will be: $vk$ (verification key), $sk$ (secret decryption key), $cd$ (ciphertext of the input data), and a proof $\pi$. In addition to data encryption, attacker $A$ may steal the data. After these actions, the attacker removes the original data and $sk$ from the victim’s machine. 
    \item \textbf{Step 3.} Attacker $A$ sends the ransom demand, its wallet account, and $vk$ to the smart contract.
    \item \textbf{Step 4/5.} The attacker's code deployed on the victim’s machine prompts the victim for the ransom. The code also computes the function $\texttt{VerifyCipherData} (\mathit{commit}, \mathit{vk}, \mathit{cd}, \pi) \rightarrow 0/1$ to convince the victim that the data can be recovered correctly.
    \item \textbf{Step 6.} The victim $V$ decides a response. If the victim is willing to pay, he/she will send a deposit of the requested number of tokens to the contract. The deposit allows the tokens to be spent only at the address of the attacker’s wallet before a timelock expires.
    \item \textbf{Step 7.} Before the timelock expires, attacker $A$ sends the correct decryption key $sk$ to contract $C$ such that $\texttt{VerifyKey}(vk, sk) = 1$. If $sk$ can be verified, contract $C$ will allows $A$ to withdraw the deposit (step 9). If attacker $A$ does not send the correct decryption key to contract $C$, after the timelock expires, the tokens will be returned to the victim.
    \item \textbf{Step 8.} The victim reads $sk$ from the contract. Using $sk$, the victim decrypts $cd$ and recovers the data by $\texttt{Dec}(sk, cd) \rightarrow data$.
\end{inparaenum}

\parait{Multi-round payment}
Besides paying the ransom to recover the data, it is possible that the attacker and victim agree on a multi-round payment plan.
In this case, the attacker promises to keep the victim’s data confidential and receives multiple payments.  The victim can discontinue the multi-round payment any time. This may occur either as response to data leakage by the attacker or simply because it is no longer necessary to keep the data private. Note that as retaliation, if the victim cancels future payment, the attacker may sell the data or disclose it to the public.

\para{Comparison with existing ransomware models}
In recent years, significant efforts have been devoted to understanding the decision-making process of ransomware attacks~\cite{ransomdecision2pay2018,li2021game,Cartwright2019ToPO,10.1007/978-3-319-94782-2_7}, particularly critical decision variables under different ransomware models. 
In the case of Ransomware 1.0, an attacker only demands cryptocurrency payment, while in Ransomware 1.5 and 2.0, the attacker may threaten to make the data public or sell the data in public. The different ransomware models affect the victim's decisions and risk mitigation strategies. For instance, the recommended practice such as having data backup and never-pay-ransom would not be optimal decision under the Ransomware 1.5 and 2.0 attack model. Empirical studies show that uncertainties involved in ransomware attacks (likelihood to recover the data after paying the ransom), trust, and protection of data confidentiality play major roles affecting a victim’ decisions and ``willingness-to-pay''.  
Compared with the previous ransomware models (1.0, 1.5, and 2.0), the new \sln{} threat brings some unique characteristics such as guaranteed data recovery using FDE and verifiable encryption, smart contract-based multi-round payment, etc. 

These characteristics can have significant impacts on the victim's decisions and extend risk management options. By applying smart contracts, \sln{} prevents either party from deviating from the agreements. It eliminates the uncertainty of data recovery after paying the ransom. As a result, it simplifies certain aspect of the ransomware response decisions. In addition, it provides more options to the victims to manage the risks with contract-based payments. For example, in the case of multi-round payment, the victim can use this option to buy additional time, which can help the victim mitigate the damage if the attacker sells the data.  For the first time, the framework considers the time dependence of the value associated with the data. Compared with the previous ransomware models, the new framework enriches risk management options for the victim. The model is a more general form where one-time ransomware payment can be treated as a special case of a scheduled payment with only one payment round.

\para{Prototyping of \slntitle{}}
To evaluate \sln{} feasibility, we have developed a prototype. 
based on tools and technologies available in the literature and the public domain, such as the implementation of VECK and FDE~\cite{atomicfairdataexchange,fde2023}, time-lock~\cite{cao2023hash}, and smart contract~\cite{vacca2021systematic}. 
In this work, we have applied FDE~\cite{fde2023} due to its efficiency compared to other verifiable encryption and fair data exchange schemes (e.g.,~\cite{saver}). 
The implementation applies CPA-secure schemes such as ElGamal or Paillier. Fair exchange supports Solidity based smart contracts. To develop a \sln{} malware prototype, the main effort is to extend and integrate the existing tools for the new scenario, ransomware data recovery instead of traditional fair data exchange. 

A typical ransomware may include many components and libraries like API hashing, API loading, handling of shadow copies, etc~\cite{9895237}. The primary focus here is on data encryption and recovery process, which is the most relevant part to the scope of this research. 
It is worth mentioning that although we have applied FDE~\cite{fde2023}, the framework is not exclusively tied with a particular design of verifiable encryption based fair data exchange. There could be a family of ransomware implementations realized by adapting other verifiable encryption schemes. 

The main overhead with the new ransomware attack scenario includes: overhead related to attacker and victim's interactions with the smart contracts, storage overhead of encrypted data, encryption and proof generation overhead by the attacker using victim's computer,  proof verification cost by the victim. In the context of FDE~\cite{fde2023}, the cost of on-chain data such as Public key and Private key uploaded to the smart contract by the attacker is constant independent from the size of the encrypted data. After paying to the smart contract, the victim retrieves the constant size Private key from the smart contract.  Instead of data bandwidth overhead in the case of FDE, in ransomware attack, the overhead is storage. Applying verifiable ElGamal encryption, it has around $11 \times$ total storage overhead with proofs included. 

\parait{Data encryption}
Applying verifiable recovery comes with additional overheads. It is not unreasonable to assume that such capability is most suited for critical data and data with high value. In fact, it is common practice for ransomware to treat data files with different encryption strategies~\cite{9895237}, for instance full encryption vs. partial encryption, only apply fully encrypting to data files below a certain size and with specific extensions (like word documents, excel data sheets, etc). 

We tested the performance of data encryption, proof generation and verification on an AMD mini-PC (AMD Ryzen 7 7840HS with 16GB DDR4 RAM (DDR4) running Ubuntu 22.04. All the experiment results are based on average of multiple iterations.
The experiment results show that it takes 239 seconds to encrypt 512KB data and generate encryption proof with ElGamal (average of 10 iterations). For 512KB encrypted data, VECK ElGamal proof generation on average requires 117.41 ms. Verifying the proof takes on average 91.036 seconds. During the experiments, execution of the code can utilize all the CPU cores, which means that the performance of encryption and proof generation can benefit from increasing number of CPU cores.

Verifiable encryption can be implemented to support symmetric ciphers~\cite{291054}. Our experiment results show that for the same data size, symmetric key based encryption and proof generation such as AES~\cite{AES_zero_knowledge_proof_circuit} have much higher overhead like running time than ElGamal based implementation. 

\parait{Smart contracts}
It is relatively straight-forward to extend the original Solidity contracts developed by~\cite{fde2023} for fair data exchange. The codes are already very efficient in terms of gas and transaction fee cost. Modifications to the contract code mainly include: When the attacker uploads the Public key, instead of demanding a single payment, the attacker can set a payment schedule. It can support more complex scenario such as asking the victim to choose between a single payment and multi-round payment option. 
Due to the gas-optimized implementation, all transactions do not incur a fee cost more than \$8.00 under a USD/ETH cost of \$2,600. 

%% file: context.tex
\section{Game Theoretical Decision Modeling}\label{sec-decision-making-model}

In this section, we develop and analyze a formal game-theoretical model tailored to \sln{}. Although game theory has been widely used to study ransomware decision-making~\cite{Cartwright2019ToPO,li2021game,10.1007/978-3-319-94782-2_7,FANG2022102685}, all the existing models typically assume uncertainty in data recovery and do not account for multi-round payment. 

\sln{} removes these assumptions by leveraging verifiable encryption and smart contracts, enabling secure data recovery and multi-round payments. This shift necessitates a new decision model to capture attacker's behavior and guide the victim toward more informed responses.

We define ``selling the data'' as the attacker monetizing the stolen data by locating potential buyers and assessing its value, while ``leaking the data'' refers to making it publicly accessible without compensation. Although these actions differ in attacker's incentives, they result in equivalent harm to the victim.


The decision process is modeled as a multi-round decision game, 
where both the attacker and the victim aim to maximize their respective utilities. Once the attacker initiates, the victim chooses whether to pay the ransom. The attacker, in turn, can either release the decryption key — allowing the first payment withdrawal while keeping the data confidential — or opt to sell or leak the data if doing so yields greater profit. Note that if the attacker does not release the key or release a wrong key, the victim will get a full refund but lose the data. 
When a payment schedule is agreed upon, the game proceeds in multiple rounds. In each round, the victim has the initiative to decide whether to continue. If the victim aborts, the attacker may act accordingly, for instance immediately make the data public or sell it. It assumes that data leakage will eventually be detected by the victim.   





Uncertainty is one of the major factors that affect the victim's decision (i.e., ``willingness to pay'' the ransom). 
Compared with the existing ransomware attacks, \sln{} alters the dynamic between the attacker and the victim:
\begin{inparaenum}[\bfseries (i)]
    \item \textit{Assurance on data recovery.} With proof-of-recovery and FDE, the uncertainty of data recovery is eliminated as a decision factor, 
    a new phenomenon never existed before.
    \item \textit{More options controlling data privacy.} The smart contract enriches payment options and enlarges the decision space for both the attacker and the victim. For instance, the contract can split the ransom into several shares and allow partial payment to the attacker over time. Although uncertainty still exists regarding data leakage, the victim has the option to cancel at any time. 
\end{inparaenum}


In the analysis, the sale price of the data may change overtime. 
This assumption applies to many real-world scenarios. 
For example, a data file may be required to be kept confidential only within a time window. After the time window expires, it is no longer required to be a secret. Examples include periodic releases of financial reports and business deals between companies that only need to remain as trade secrets before they are made public. 

\subsection{ Multi-round payment model }


\tablename~\ref{tab-notation} lists the notations in the following analysis. There are $n$ payment rounds. At the beginning of the first round, the victim chooses whether to pay the first ransom or not. If the victim pays the first ransom, the attacker will return to decryption key with probability $\beta_r(0\le\beta_r\le 1)$. Note that $zkRansomware$ ensures that the attacker can withdraw the payment only if the key is correct. Otherwise, the victim will be refunded. 
At the beginning of subsequent round $i(i\in\{1,2,\dots,n\})$, the victim is asked to pay the $i$-th ransom. The victim will decide whether to continue or abort the game. If the victim makes the payment as required, the attacker then decides whether to maintain confidentiality of the data or to sell it with a probability of $\beta_i$ after accepting the ransom. This means that for $i$-th round, the probability for the attacker to keep the victim's data confidential will be $1-\beta_i$. On the other hand, if the victim discontinues, the attacker will simultaneously proceed to either sell or leak the data, which will effective terminate the game. This process will continue until either reaching the final stage or termination by the victim. 
 
{\scriptsize
\begin{table}
\vspace{-0.2in}
\scriptsize
\centering
\caption{Notations used in the analysis.}
\vspace{-0.1in}
\label{tab-notation}
\begin{tabular}{p{0.5in}p{1.9in}|p{0.5in}p{1.7in}}
    \toprule 
    Notation& Description &Notation& Description\\
    \midrule
      $U_{a,i}$   &  the utility of attacker from round $i$ &
      $U_{v,i}$ & the utility of victims from round $i$ \\
      
      $p_i$   &  payment decision, equals to 1 when paying the ransom in round i or equals to 0 &
      $s_i$ & selling decision, equals to 1 when selling the data or equals to 0\\
      
      $C_r$ & cost to recover the data&
      $A_i$ & selling profit in round $i$ \\
      
      $R_i$  & ransom amount in round $i$  &
       $V$  & victim's data value \\
      
      $L_i$ & loss from data leakage in round $i$ &
       $n$  & number of payment rounds\\
       $\beta_r$ &  probability of file returning in the first round &
       $\beta_i$ &  probability of keeping data confidential in the $i$-th round    \\
    \bottomrule
\end{tabular}
\vspace{-0.2in}
\end{table}
}


There are some assumptions worth mentioning before further analysis. First the moment the attacker sells the victim's data is assumed to occur immediately after the victim's decision on stopping payment. The value of the data may change over time for instance decrease overtime or reduce to zero if it is unnecessary to keep it confidential after a timeline, and consequently, the profit from selling the data likely diminishes over time. Therefore, if the attacker decides to sell the data, it is optimal to do so as early as possible. 
On the victim's side, we assume that the damage caused by data leakage or loss of privacy will be eventually observed by the victim, and as a result, the victim will abandon any subsequent ransom payments immediately after discovering the leakage. 

We first consider the scenario where the attacker has no credibility (the worst reputation which corresponds to $(\beta_r,\beta_1,\dots,\beta_n)=(0,1,1,\dots,1)$) and makes decisions solely to maximize profit. In this case, we summarize the strategies taken by the attacker and victims in the following theorem.

\begin{The}\label{th1}
    Faced with the worst-reputation attacker, victims will never pay the ransom. The attacker will choose to sell the data.
\end{The}
The proof can be found in Appendix \ref{appendix-proofs} $\square$.

When facing a completely non-credible attacker, even a multi-round ransom payment scheme fails to increase the victim's willingness to pay. This is because, in the final stage, the attacker will invariably choose to sell the data to seek greater profit, which leads the victim to be unwilling to pay the ransom for that last stage. Consequently, the situation in the penultimate stage degenerates to that of the final stage. By backward induction, and as the reasoning proceeds to the first stage, the attacker would also refuse to recover the victim's data in pursuit of higher profits, resulting in the victim's persistent unwillingness to pay any ransom at all. Note that in this case, the analysis indicates that it would be a waste of time for the attacker to even raise the ransom demand because due to the design of \sln{}, it is not an option for the attacker to withdraw the ransom without releasing the correct key to the victim. 

To demonstrate the necessity of maintaining better reputation, we will next analyze the victim's decision-making and the attacker's payoff under the scenario of perfect reputation in the next theorem which means when victims pay the ransom, the attacker will return the key and keep the data confidential(corresponds to the case $(\beta_r,\beta_1,\dots,\beta_n)=(1,0,0,\dotsm0)$). 

It is important to point out that the attacker won't gain financial benefit from creating many identities. In the eye of the victim, a fresh attacker without history, is equivalent with the attacker having the lowest reputation. This suggests that for any rational attacker, creating many identities wont be an optimal strategy to gain higher ransom profit.   

Before our analysis, we recall the definition of an indicator function $I_{A}(x)$ which equals to 1 when $x\in A$ or it equals to 0.
\begin{The}\label{th2}
    We can derive the victim's strategy through the following steps when facing a perfect reputation attacker. We assume $a_n=\min\{R_n, L_n\}, a_{n-1}=\min\{R_{n-1}+a_n,L_{n-1}\},\dots,a_i=\min\{R_i+a_{i+1},L_i\}$ where $i\in\{2,\dots,n-1\}$ and $a_1=\min\{R_1-V+a_2 ,L_1\}$. Then victims will choose to abort the game in round t.
    \begin{equation}
        t=\min_{I_{\{L_j\}}(a_j)=1} j 
    \end{equation}
\end{The}
The proof can be found in Appendix \ref{appendix-proofs} $\square$.

Different from the worst reputation case, there will be victims who choose to pay the ransom facing an attacker with  perfect reputation. Apparently, better reputation and \sln{ }improve the ``willingness to pay''. Meanwhile, continuation of receiving payments may delay data sale by the attacker. Maintaining a better reputation is a tradeoff between obtaining additional profit from data sale and gaining more profit from the victim. 

\begin{figure}
     \vspace{-0.3in}
     \centering    
     \tiny  
     \begin{minipage}[t]{0.6\linewidth}
         \centering
         \includegraphics[width=\linewidth]{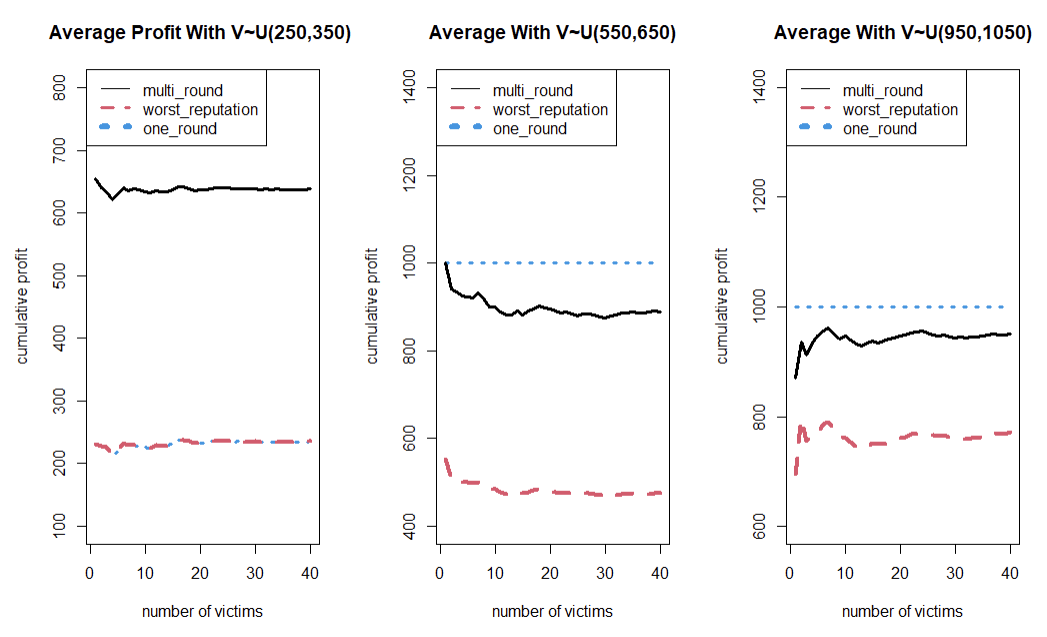}
         \vspace{-0.1in}
         \caption{Average attacker profit under different data values.}
         \label{fig:average_profit}
     \end{minipage}    
    
     \begin{minipage}[t]{0.6\linewidth}
         \centering
         \includegraphics[width=\linewidth]{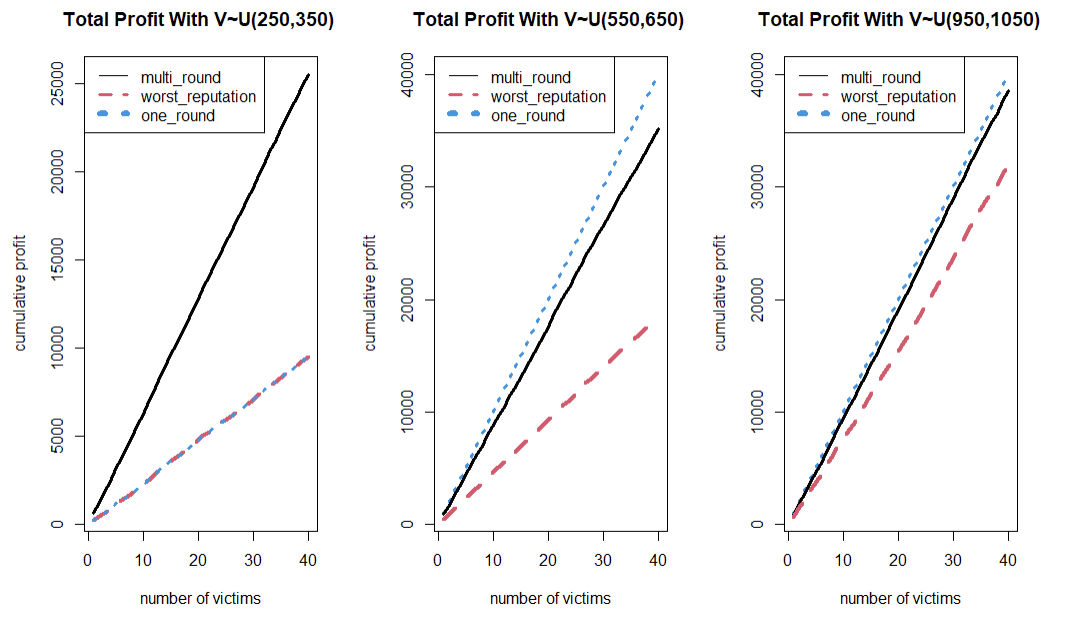}
         \vspace{-0.1in}
         \caption{Total attacker profit under different data values.}
         \label{fig:total_profit}
     \end{minipage}
         \hfill
     \begin{minipage}[t]{0.35\linewidth}
         \centering
         \includegraphics[width=\linewidth,height=1.6in]{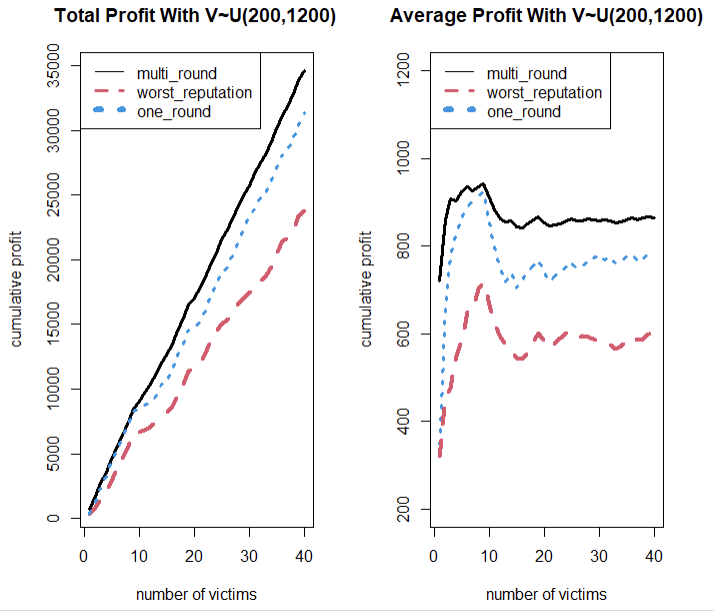}
         \vspace{-0.1in}
         \caption{Profit under different data value ranges.}
         \label{fig:data_values}
     \end{minipage}
     \vspace{-0.1in}
\end{figure}

In order to show the necessity of maintaining a good reputation, we show the simulation result in Figure ~\ref{fig:average_profit}. It considers 3 cases: one-round payment, multi-round payment with perfect reputation, and the worst reputation attacker case.  



The simulation considers a total of 8 payment rounds, with 40 victims. The ransom demanded from each victim is 1,000. In single-round payment, the entire ransom is paid in a lump sum, while in multi-round payment, a relatively larger portion of the ransom is allocated to the first payment, with the remaining amount distributed evenly across the subsequent rounds. This setup is justified. From (Theorem 2), we can observe that in the first payment, since the victims can directly recover their data, they are willing to bear a relatively higher ransom. We examine three different data value distributions, each following U(250, 350), U(550, 650), and U(950, 1050), respectively, where U(a, b) represents a uniform distribution ranging from a to b. These correspond to three scenarios: over charged by the attacker, reasonably charged, and relatively low charged. The simulation results indicate that multi-round payment yields higher profit for the attacker than the other two scenarios when the victims are overcharged. 

In the multi-round setup, the victims consider paying the initial portions of the ransom to recover the data, and once the utility of keeping the data confidential declines to a certain level, the victims will stop the payment.  When the data value is large, the victims are in general willing to pay a higher ransom. 
For data with high value, although its value may decline over time, there is likely need to keep the data confidential, 
which narrows the profit gap between multi-round and single-round ransom payment.

Finally, we consider a scenario with mixed data values, where the data values follow a uniform distribution between 200 and 1,200. The results are in Figure ~\ref{fig:data_values}. In this case, the attacker clearly can gain higher profits under the multi-round payment scheme. 
The result may be counter-intuitive because under a composite distribution of data values, and overcharging scenario, the victims have the freedom and option to discontinue any time. The results based on game theoretical modeling show that the potential profit to the attacker from the multi-round payment scheme still surpasses the profit of single-round payment.

\subsection{Multi-round payment with imperfect reputation}

From the previous analysis, one can observe that maintaining a certain reputation is necessary for the attacker for higher profit. Below, we will consider an imperfect reputation model. In the $i$-th round, the attacker will sell the data with probability $\beta_i$ and keep the data confidential with probability $1-\beta_i$. 
We will represent the attacker's reputation with the vector $(\beta_r,\beta_1,\dots,\beta_n)$. The first theorem below provides the victim's decision in each round given the attacker's known reputation.

\begin{The}\label{th3} 
    One can derive the victim' decision strategy through the following when facing an attacker whose reputation is $(\beta_r,\beta_1,\dots,\beta_n)$. We assume $b_n=\min\{R_n+\beta_nL_n,L_n\},b_{n-1}=\min\{R_{n-1}+\beta_{n-1}L_{n-1}+(1-\beta_{n-1})b_n,L_{n-1}\},\dots,b_{i}=\min\{R_{i}+\beta_{i}L_{i}+(1-\beta_{i})b_{i+1},L_{i}\}$ where $i\in\{2,3,\dots,n-1\}$ and $b_1=\min\{R_1+(1-\beta_r)L_1+\beta_r(-V+\beta_1L_1+(1-\beta_1)b_{2}),L_1\}$. Then victim pays the first t-1 ransoms and chooses to discontinue in round t if and only if the data has not been leaked or sold before time t and $I_{\{L_l\}}(b_l)=0$ for $l=1,2,,\dots,t-1$,$I_{\{L_t\}}(b_t)=1$.
\end{The}
The proof can be found in Appendix \ref{appendix-proofs} $\square$.

Given the victim's decision in each round, we can determine the optimal reputation that the attacker must maintain to maximize its expected payoff when the value of the victim's data is fixed. We will present the method for solving for this optimal reputation in the following theorem.

\begin{The}\label{th4}
    With n payment rounds, the optimal reputation for the attacker can be determined by solving n linear programming equations. Let $(x_1^{(i)},x_2^{(i)},\dots,x_{i+1}^{(i)})$ be the optimal point for the i-th optimization problem. After deriving the optimal reputation, one can obtain the expected attacker payoff.
\end{The}  

Due to page limit, we skip the detailed equations. The proof can be found in Appendix \ref{appendix-proofs} $\square$.

We partition the domain of reputation into $n+1$ distinct regions based on the different strategies adopted by the victim in response to the attacker’s reputation, as given in Theorem 3. These regions respectively correspond to the victim's decision of paying no ransom, paying only the first ransom, paying the first two ransoms, …, up to paying all the $n$ round‑wise ransoms. In each region, the expression for the attacker’s expected payoff differs.  Optimal reputation can be solved separately in each region with the goal of maximizing the expected payoff, then compare the payoffs corresponding to these optimal reputations. Finally, one can obtain the globally optimal reputation for the attacker and the optimal strategy that the victim adopts under that reputation. 

\subsection{Overcharging attacker}

In this section, we will demonstrate that for the attacker who over-demands ransom, a multi-round payment mechanism can yield significantly higher expected total payoff compared to one-time ransom payment. Let the total ransom amount be R, which may exceed $V + L$. Let us denote it as $NV + L_1$ ($N > 1$). We first consider the victim's decision in a single-round scenario. In this case, the ransom amount is R, 
\begin{Pro}
    With only one payment round, the victim will choose to pay if and only if $R<\beta_r(V+(1-\beta_1)L_1)$.
\end{Pro}\label{th5}

One can simply obtain the proposition above through the expected utility of the victim with one payment round, that is $EU_{v,1} = p( -R-(1-\beta_r)L_1+\beta_r(V-\beta_1L) ) - (1-p)L_1$. When the ransom amount far exceeds V + L, even under the best reputation condition (i.e. $\beta_r=1,1-\beta_1=0$ ), in a single-round ransom payment scenario, the victim will inevitably choose to refuse paying. For the attacker, the resulting payoff can only be $A_1$, which represents the only profit from immediate sale of the data.  

We then turn to the multi-round payment case. We observe that not every ransom arrangement can achieve an expected return exceeding $A_1$. As a simple example, if the victim is required to pay almost the entire ransom in the first round, with only a small portion left for the remaining rounds, then this multi-round payment scheme essentially degenerates to a single-round ransom demand. When the ransom amount is large, the victim will inevitably choose to refuse payment from the very beginning. To incentivize the victim to pay the ransom and thereby increase the total expected return, the attacker must adopt a reasonable ransom arrangement. Such a reasonable ransom arrangement is quite easy for the attacker to work out. For instance, setting the first-round ransom to $V$ and evenly distributing the remaining amount across the subsequent rounds would suffice. Actually if the first ransom is small than $V$, which we assume is $(1-\gamma)V\;(\gamma>0)$, victim will pay the first ransom with a corresponding reputation setting. However the larger $\gamma$ is, the less profit the attacker will gain. In terms of the attacker's profit, there is no need to set $\gamma$ too small. Under the corresponding optimal reputation conditions, one can show that the victim will at least pay the first installment of the ransom. Furthermore, one can estimate the lower bound of the attacker’s expected profit under this reasonable ransom arrangement.

\begin{The}\label{th6}
    There are a total of n payment rounds, with the total ransom amount being $R = NV + L_1$ $(N > 1)$. We assume that the ransom amount for the first round is $(1-\gamma)V$ where $\gamma>0$ and $(1-\gamma)V+A_2>A_1$. Under the corresponding optimal reputation conditions, let $Pro$ be the attacker profit, then it satisfies: $Pro>(1-\gamma)V+A_1-C_r$. 
\end{The}
The proof can be found in Appendix \ref{appendix-proofs} $\square$.

The theorem above indicates that when the total ransom amount is relatively high, setting the initial ransom payment slightly below V results in a higher expected profit. In fact, even if the initial ransom amount is set somewhat higher, as long as the attacker maintains a good reputation, the victim may still choose to pay at least the first ransom. This allows the attacker to further increase the expected return. Regarding another aspect of the aforementioned theorem, when deriving the lower bound for the expected profit of an attacker who over-demands ransom, we only consider the scenario where the attacker receives only the first ransom payment. This consideration is relatively intuitive because the total ransom amount is substantial. Based on the analysis from the previous sections, to incentivize the victim to continue paying the ransom, the first ransom can generally be set higher since the victim can immediately recover the data. In contrast, the subsequent ransom amounts are constrained by the diminishing marginal value of the recovered data and cannot be set too high when its future value is uncertain. Therefore, distributing the remaining (N-1)V+L ransom across the remaining n-1 rounds would likely reduce the victim's ``willingness to pay" of the future ransom. Consequently, in such a scenario, it is reasonable to consider only the first ransom payment when determining the corresponding lower bound for the expected return. Compared to a single round of ransom payment, the expected return in multi-round situations is at least increased by $(1-\gamma)V-C_r$. 

%% file: simulations.tex
\section{Simulation Experiments}

In this section, we provide simulation experiment results of the ransomware attack using the optimal reputation decision model derived previously. The study interprets the results from the following perspectives: (i) configuration of the optimal reputation; (ii) simulated profit under the optimal reputation.

\subsection{Optimal Reputation}

Before conducting the simulation, we first present the simulation setting. Without losing generality, let $R$ be the amount of tokens that the attacker demands from the victim, payable over $n=6$ rounds. The first ransom payment accounts for $50\%$ of the total ransom, with the remaining amount evenly distributed across each subsequent round. The data value is assumed to be $V$. In the $i$-th payment round, if data leakage occurs, the resulting loss is $L_i=f(\frac{i}{n})V$. We consider three different functional forms for f, corresponding to three distinct declining trends: $y=(1-x)^2, y=1-x$, and $y=\sqrt{1-x^2}$. The subsequent simulations consider three patterns of data value decline that occur randomly with equal probability. Further, let $A_i=0.7V_i$ be the profit through data sales for the attacker in round $i$. We first consider the optimal reputation under different scenarios. Let 500 be the data value. 




\begin{figure}
    \vspace{-0.3in}
    \centering    
     \tiny  
     \begin{minipage}[t]{0.45\linewidth}
         \centering
         \includegraphics[width=\linewidth]{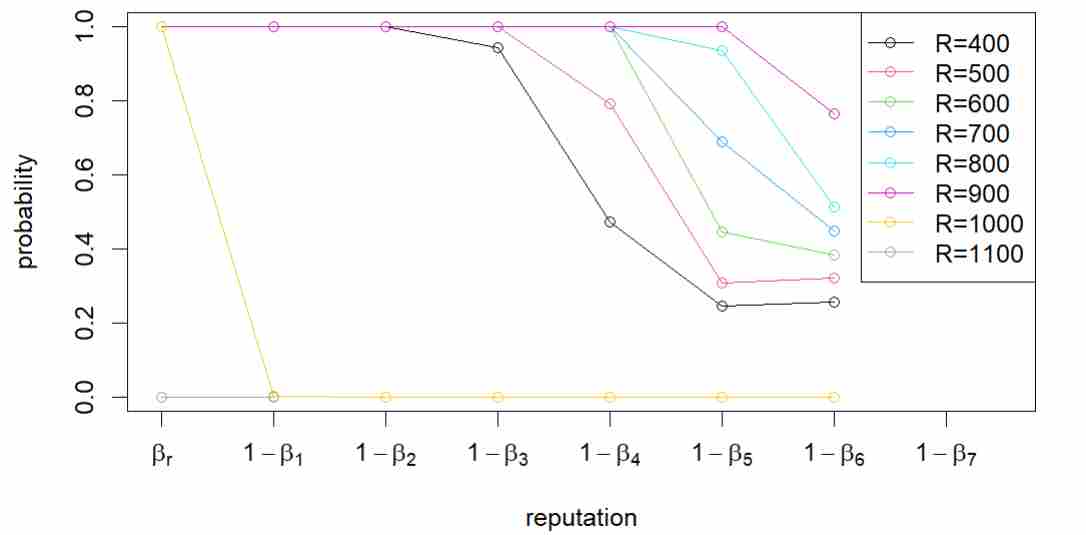}
         \vspace{-0.1in}
         \caption{Optimal reputation with $f(x) = (1-x)^2 $.}
         \label{fig:convex}
     \end{minipage}    
     \hspace{2pt}
     \begin{minipage}[t]{0.45\linewidth}
         \centering
         \includegraphics[width=\linewidth]{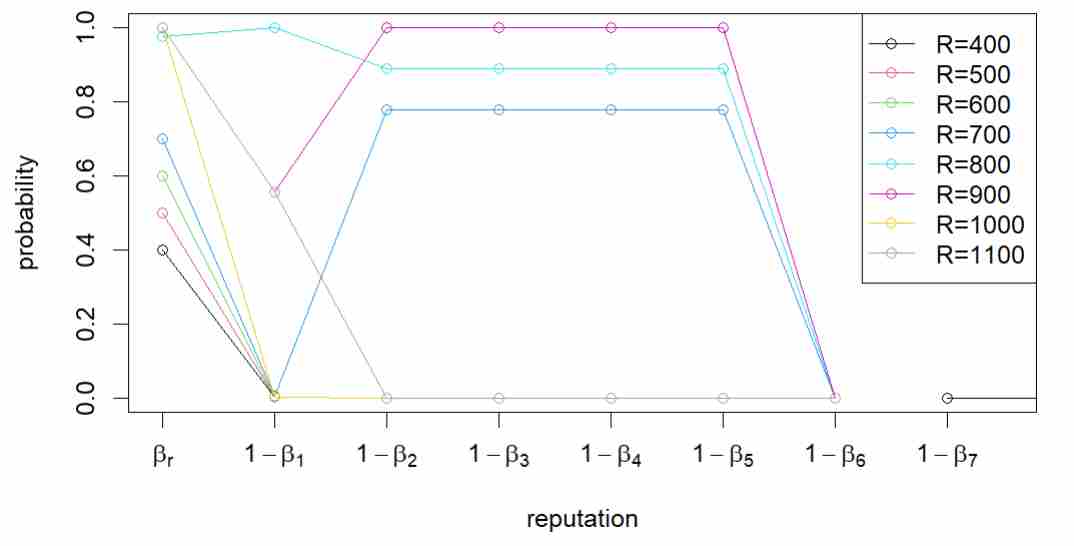}
         \vspace{-0.1in}
         \caption{Optimal reputation with $f(x) = 1-x $.}
         \label{fig:uniform}
     \end{minipage}
    
     \begin{minipage}[t]{0.45\linewidth}
         \centering
         \includegraphics[width=\linewidth]{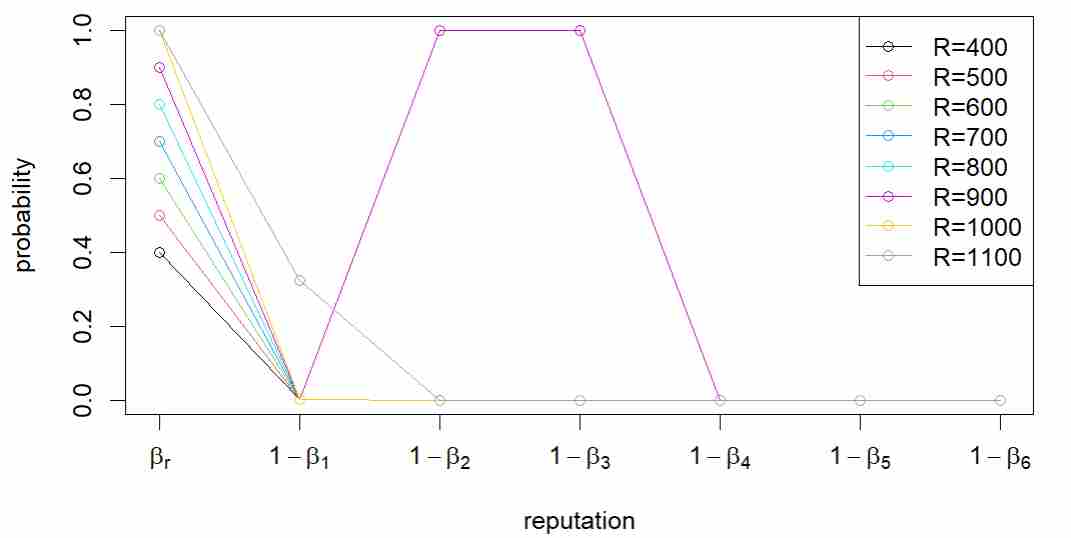}
         \vspace{-0.1in}
         \caption{Optimal reputation with $f(x) = \sqrt{1-x^2} $.}
         \label{fig:concave}
     \end{minipage}
     \hspace{2pt}
     \begin{minipage}[t]{0.45\linewidth}
         \centering
         \includegraphics[width=\linewidth]{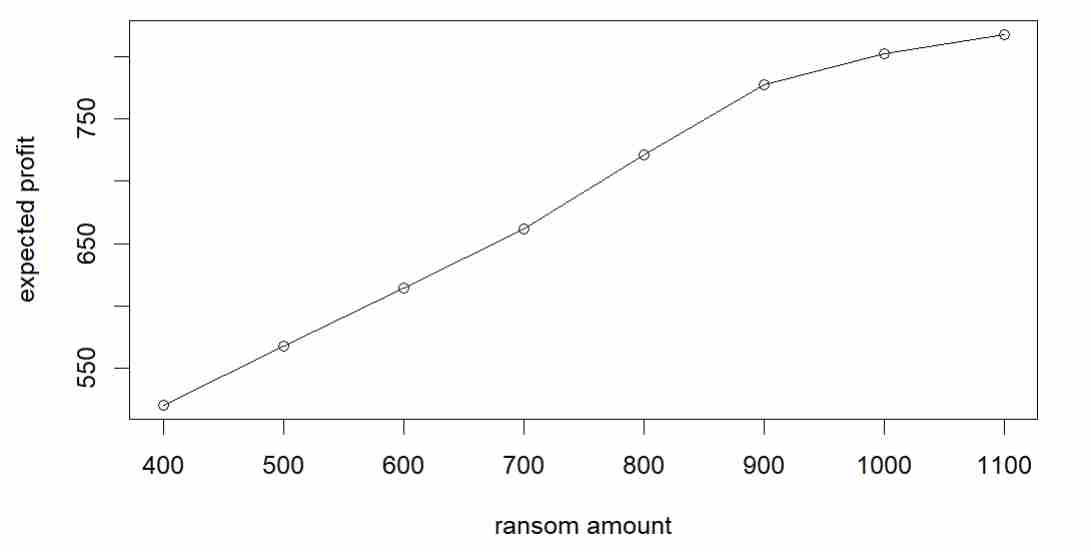}
         \vspace{-0.1in}
         \caption{Expected profit under different ransom amounts.}
         \label{fig:expected_profit}
     \end{minipage}
    \vspace{-0.2in}
 \end{figure}

To make the figures more intuitive, the horizontal axis represents the probability of returning the data after receiving the first ransom, as well as the probability of maintaining data confidentiality in each subsequent round. In Figure \ref{fig:convex}, one can observe that when the data value decays more slowly in the initial rounds, the attacker tends to maintain data confidentiality to secure subsequent ransom payments. In Figures \ref{fig:uniform} and \ref{fig:concave}, where the data value decays rapidly in the early rounds, the attacker is aware that the victim is less likely to pay the ransom when the remaining data value is low. Therefore, the attacker is more inclined to sell the data early on. For this scenario, the data value is set at 500. The probability of returning the data after receiving the first ransom remains near 1 when the ransom amount exceeds 800. When the ransom amount is relatively small, simply receiving the ransom is not sufficiently profitable for the attacker. Thus, the attacker may choose not to return the key with a certain probability and instead sell it to increase gains, while the victim continues to pay the ransom. In most cases, the attacker demands a ransom far exceeding the data value. Therefore, maintaining a high probability of returning the key is not disadvantageous for the attacker. The decision game ensures that returning the key after receiving the ransom is beneficial for the attacker, as it helps maintain an optimal reputation and enhance the overall profit.

\subsection{Profit Analysis}
When demonstrating the advantage of the optimal reputation in terms of attainable profit, we simply selected the optimal reputation under the condition where R=800 and the data loss coefficient decreases according to the function y=1-x. First, we consider the expected profit that the attacker can obtain under different ransom amounts. In Figure \ref{fig:expected_profit}, it can be observed that as the ransom amount increases, the expected profit also increases and gradually approaches the lower bound according to Theorem \ref{th6}. 

\begin{figure}
    \vspace{-0.3in}
    \centering    
    \tiny  
    \begin{minipage}[t]{0.7\linewidth}
        \centering
        \includegraphics[width=\linewidth,height=1.6in]{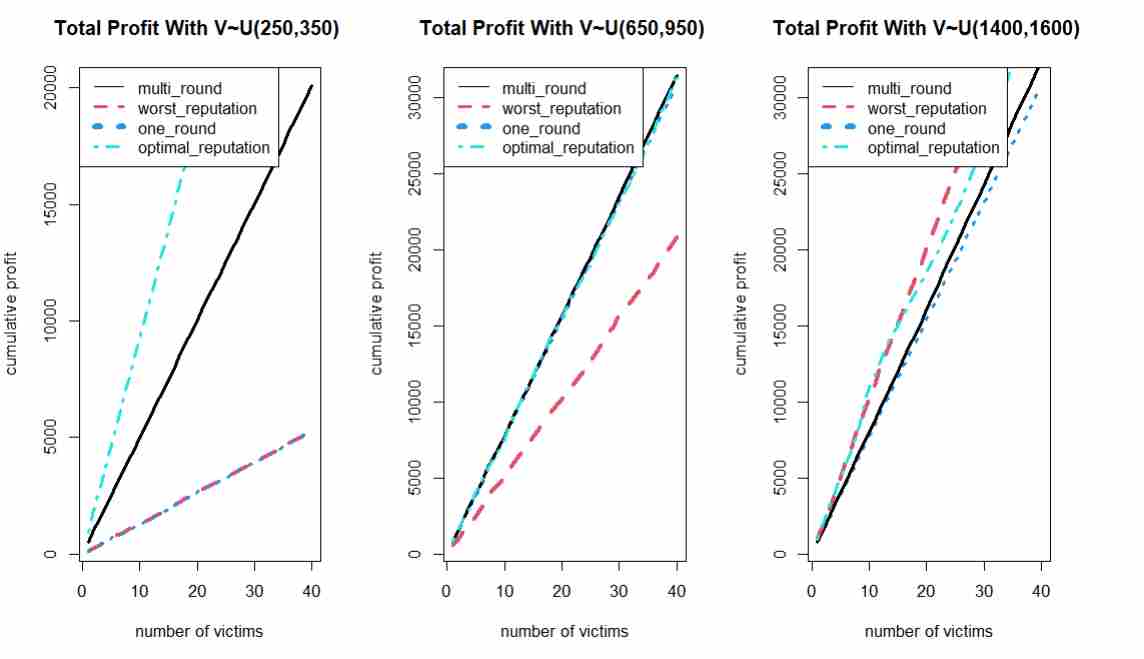}
        \vspace{-0.1in}
        \caption{Simulated profit under different data values and reputations.}
        \label{fig:sim_proft1}
    \end{minipage}    

    \begin{minipage}[t]{0.7\linewidth}
        \centering
        \includegraphics[width=\linewidth,height=1.6in]{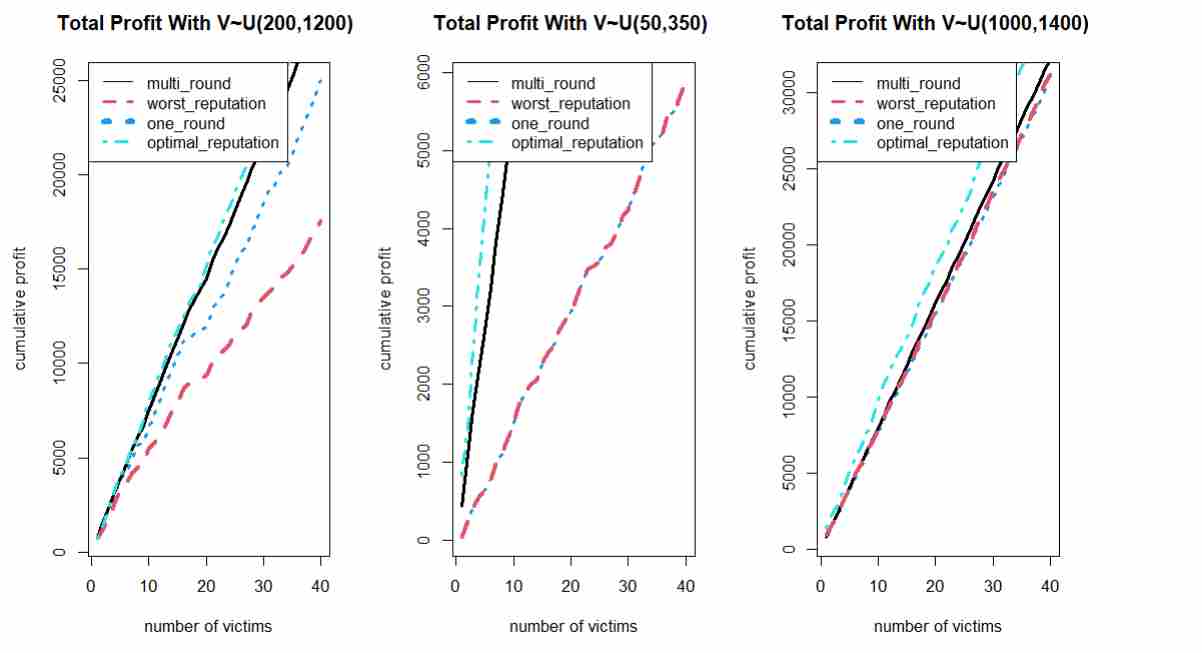}
        \vspace{-0.1in}
        \caption{simulated profit under different data value ranges.}
        \label{fig:sim_proft2}
    \end{minipage}
        \vspace{-0.3in}

\end{figure}

Next, we will compare the simulated profits under four different scenarios: the worst-case reputation, perfect reputation in a single round, perfect reputation in multiple rounds, and optimal reputation in multiple rounds. Under each reputation scenario, we consider the cases of excessive ransom demand (where the data value V follows a uniform distribution $U(250, 350)$), moderate ransom demand (V follows $U(650, 950)$), and relatively low ransom demand (V follows $U(1000, 1300)$) in Figure \ref{fig:sim_proft1}. The optimal reputation typically lies between the worst-case reputation and the perfect reputation. Intuitively, it tends to perform well when the perfect reputation under-performs. In the first subplot of Figure \ref{fig:sim_proft1}, when the attacker demands an excessive ransom, even a perfect reputation case cannot guarantee that the victim will pay the full ransom. In most cases, the victim pays only the initial few ransom installments and refuses to continue with the subsequent payments. In such scenarios, if the attacker's reputation deteriorates slightly compared to the perfect reputation case, the attacker may choose to sell the data earlier. Although the victim still pays only the initial installments, the early sale of the data results in increased overall profit. If the demanded ransom is significantly lower than the data value, the worst-case reputation (which involves selling the data directly) performs better instead. This is because, even under such a reputation, the victim may still pay the ransom. Similarly, the slightly worse reputation (optimal reputation) yields higher returns compared to the perfect reputation case. In Figure \ref{fig:sim_proft2}, under both extremely high and extremely low ransom demands, the optimal reputation yields higher profits. However, when the attacker has only a broad estimate of the victim’s data value — as illustrated in the figure where V follows a $U(200, 1200)$ distribution — the profits gained under the optimal reputation and the perfect reputation case are quite similar. Therefore, when the estimation of the data value is relatively accurate, the optimal reputation case tends to deliver higher profit.

%% file: mitigation.tex
\section{Mitigation Strategies}
According to both theoretical analysis and simulation experiment, \sln{} is attractive to the attacker under certain situations because of higher expected profit and increased ``willingness-to-pay'' by the victim. Despite its overhead, \sln{} could be used in conjunction with the traditional attack approach where \sln{} is selectively applied to the most valuable data. Regarding mitigation strategies, our theoretical model provides a decision support tool for understanding the attacker's behavior and decisions. For detecting \sln{}, verifiable encryption likely has unique behavior profiles such as using certain types of operations and instructions to accelerate computation, which can be leveraged for detection. Last but not the least, \sln{} highlights the importance to protect data privacy at rest with approach such as encrypted file systems or disk volumes.

%% file: conclusion.tex
\section{Conclusion}

This paper presents an emerging ransomware threat that applies verifiable encryption, smart contracts, and multi-round cryoto-payment.  Compared with the previous ransomware models (1.0, 1.5 and 2.0), this new threat exhibits unique decision making dynamics between the attacker and the victim.
It might seem counterintuitive, the new attack scenario expands the decision options for the victim managing data recovery and the risk of losing data privacy.
Both game-theoretical analysis and simulation study suggest that there exists new decision equilibrium between the attacker and the victim, distinguishing from all the known ransomware decision models.

%% file: appendix.tex
\appendix

\section{Proofs of Major Theorems}\label{appendix-proofs}

\subsection{Proof of Theorem~\ref{th1}}

\begin{proof}
    There are n ransom payment rounds. At the beginning of ith round, the victim decides whether to pay the ransom $R_i$. After that, the attacker decides whether to sell(leak) the data or not. Both the attacker and the victim's strategies can be obtained through backward induction. The utilities for the attacker and the victim in the last round n are: 
    \begin{equation}
        \begin{split}
            U_{a,n} &= p_n(R_n+s_nA_n) +(1-p_n)A_n\\
            U_{v,n} &= p_n(-R_n - s_nL_n) - (1-p_n)L_n \\
        \end{split}
    \end{equation}
    We know only the case when $(s_n,p_n)=(1,0)$ is the Nash equilibrium from the following chart. 
    
\begin{tabular}{|c|c|c|}
  \hline
  \diagbox{$p_n$}{$s_n$} & 1 & 0  \\
  \hline
   1 & ($R_n + A_n, -R_n-L_n$) & ( $R_n, -R_n$ )  \\
  \hline
   0 & ( $ A_n,-L_n $ ) & ( $A_n, -L_n$ )  \\ 
  \hline
\end{tabular}\\
Then the utility of attackers is $A_n$ and the utility of victims is $-L_n$. We then turn to the (n-1)th payment round.
\begin{equation}
    \begin{split}
         U_{a,n-1} &= p_{n-1}(R_{n-1}+s_{n-1}A_{n-1}+(1-s_{n-1})U_{a,n}) +(1-p_{n-1})A_{n-1}\\
         &=p_{n-1}(R_{n-1}+s_{n-1}A_{n-1}+(1-s_{n-1})A_{n}) + (1-p_{n-1})A_{n-1} \\
        U_{v,n-1} &= p_{n-1}(-R_{n-1} - s_{n-1}L_{n-1}+(1-s_{n-1})U_{v,n}) - (1-p_{n-1})L_{n-1} \\
        &=p_{n-1}(-R_{n-1} - s_{n-1}L_{n-1}-(1-s_{n-1})L_n) - (1-p_{n-1})L_{n-1}\\
    \end{split}
\end{equation}
Draw the utility chart in payment round (n-1) and obtain the Nash equilibrium.\\
\begin{tabular}{|c|c|c|}
  \hline
  \diagbox{$p_{n-1}$}{$s_{n-1}$} & 1 & 0  \\
  \hline
   1 & ($R_{n-1}+A_{n-1}, -R_{n-1}-L_{n-1} $) & ( $R_{n-1}+A_{n}, -R_{n-1}-L_n$ )  \\
  \hline
   0 & ( $ A_{n-1},-L_{n-1} $ ) & ( $A_{n-1}, -L_{n-1}$ )  \\ 
  \hline
\end{tabular}\\
The strategy taken by attackers and victims are $(s_{n-1},p_{n-1})=(1,0)$ which is the same as the strategy in round $n$. Similarly one can replace round number $n$ by $i+1$, and obtain the strategies chosen in round $i$. Through backward induction, one can learn that when $i \ge 2$, the victim always refuses to pay the ransom and the attacker always sells the data. The corresponding utility in round i is $(U_{a,i},U_{v,i})=(A_i,-L_i)$.   We then consider the first round.
\begin{equation}
        \begin{split}
            U_{a,1} &= p_1(R_1-rC_r+s_1A_1+(1-s_1)U_{a,2}) +(1-p_1)A_1\\
            &=p_1(R_1-rC_r+s_1A_1+(1-s_1)A_2) +(1-p_1)A_1 \\
            U_{v,1} &= p_1(-R_1 + rV - s_1L_1+(1-s_1)U_{v,2}) - (1-p_1)L_1 \\
            &= p_1(-R_1 + rV - s_1L_1-(1-s_1)L_2) - (1-p_1)L_1
        \end{split}
\end{equation}
One can learn from the utility of the attacker that even if the victim pay the ransom, the most profitable strategy is accepting the ransom and then sell the data at once. The only Nash equilibrium is that the victim doesn't pay the ransom and the attacker will sell the data at once.    
\end{proof}

\subsection{Proof of Theorem~\ref{th2}}

\begin{proof}
    Let's analyze the strategies taken by the victim. First, apply backward induction and write the victim's utility in round n.
    \begin{equation}
        U_{v,n} = -p_nR_n-(1-p_n)L_n
    \end{equation}
    Obviously the victim chooses to pay if $R_n<L_n$ with the utility $U_{v,n}=-R_n$. Or refuses to pay if $R_n\ge L_n$ with the utility $U_{v,n}=-L_n$. The victim's utility in final round could be written as $U_{v,n}=-\min\{R_n,L_n\}$=-$a_n$. We then consider the victim's choices in round n-1. 
    \begin{equation}
        U_{v,n-1} = p_n(-R_{n-1}-a_n)+(1-p_n)(-L_n)
    \end{equation}
    With similar analysis, the victim's utility in round n-1 is $U_{v,n-1}=-\min\{R_{n-1}+a_n,L_{n-1}\}$. The victim will pay the ransom in round n-1 if $R_{n-1}+a_n<L_{n-1}$ otherwise discontinue. Using the same methodology, one can calculate the attacker's utility and optimal decision in i-th payment round where $2\le i\le n-1$. $U_{v,i} = -\min\{R_{i}+a_{i+1},L_i\}$. If $R_i+a_{i+1}>L_i$, the victim discontinues in  round i. If $R_i+a_{i+1}<L_i$, the victim chooses to pay in round i. In the first round, one can derive victims' utility. 
    \begin{equation}
        U_{v,1} = p_{1}(-R_1+V+U_{v,2})-(1-p_1)L_1
    \end{equation}
    where $U_{v,2}=\min\{R_2+a_3,L_2\}$. Therefore the victim will pay until round t where t is the earliest time to discontinue, which makes the victim's utility equal to the leakage loss. Then the expression of round t is that of $t=\min_{I_{\{L_j\}}(a_j)=1} j $
\end{proof}

\subsection{Proof of Theorem~\ref{th3}}

\begin{proof}
We analyze the strategies taken by the victim. Through backward induction, one can write the victim's utility for final round n.
    \begin{equation}
        U_{v,n} = -p_n(R_n+\beta_nL_n)-(1-p_n)L_n
    \end{equation}
    Obviously victims choose to pay the ransom if $R_n<(1-\beta_n)L_n$ with the utility $U_{v,n}=-R_n-\beta_nL_n$. They refuse to pay the ransom if $R_n\ge (1-\beta_n)L_n$ with the utility $U_{v,n}=-L_n$. The victim's utility in round n could be written as $U_{v,n}=-\min\{R_n,L_n\}$=-$b_n$. We then consider the victim's choices in round n-1. 
    \begin{equation}
        U_{v,n-1} = p_{n-1}(-R_{n-1}-\beta_{n-1}L_{n-1}-(1-\beta_{n-1})b_n)+(1-p_{n-1})(-L_{n-1})
    \end{equation}
    With similar analysis, the victim's utility in round n-1 is $U_{v,n-1}=-\min\{R_{n-1}+\beta_{n-1}L_{n-1}+(1-\beta_{n-1})b_n,L_{n-1}\}$. The victim will pay the ransom in round n-1 if $R_{n-1}+(1-\beta_{n-1})(b_n-L_{n-1})<0$ or discontinue. Using the same methodology, one can calculate the attacker's utility and optimal decision in $i$-th payment round where $2\le i\le n-1$. $U_{v,i} = -\min\{R_{i}+\beta_iL_i+(1-\beta_i)b_{i+1},L_i\}$. If $R_i+(1-\beta_i)(b_{i+1}-L_{i})>0$, the victim discontinue in round i. Otherwise the victim chooses to pay in round $i$. For the first payment round, one can obtain the victim's utility. 
    \begin{equation}
        U_{v,1} = p_{1}[-R_1-(1-\beta_r)L_1+\beta_r(V-\beta_1L_1+(1-\beta_1)U_{v,2})]-(1-p_1)L_1
    \end{equation}
    where $U_{v,2}=-b_2$. In this case, To summarize the condition which makes the victim refuse paying the ransom. The victim refuses to pay in the $i$-th round if and only if the attacker keeps the data confidential for the first $i-1$ rounds but leaks the data or sells the data in the round $i$ or in the first $i-1$ rounds. For the victim, paying the ransom always results in less loss but paying the $i$-th ransom leads to more loss than refusing to pay which means $ \min_{I_{\{L_j\}}(a_j)=1} i$.
\end{proof}

\subsection{Proof of Theorem~\ref{th4}}

\begin{proof}
First consider all possible reputation settings under which the victim will pay the ransoms for the first $i$ rounds and refuse to pay in the $(i+1)$-th round where $i\ge2$. The reputation $(\beta_r,\beta_1,\dots,\beta_n)$ needs to satisfy:
\begin{equation}\label{inequalty_con_ofiransom}
\begin{cases}
R_i+(1-\beta_i)(L_{i+1}-L_i) < 0 \\
R_{i-1}  + (1-\beta_{i-1})(R_i-L_{i-1}+\beta_iL_i+(1-\beta_i)L_{i+1}) <0 \\
\dots\\
R_{i-k} + (1-\beta_{i-k})( R_{i-k+1} - L_{i-k}+L_{i-k+1})+\\(1-\beta_{i-k})(1-\beta_{i-k-1})( R_{i-k+2} - L_{i-k+1}+L_{i-k+2})+\dots+\\ \prod_{l=i-k}^{i-1}(1-\beta_l)(R_i-L_{i-1}+L_i)+\prod_{l=i-k}^{i}(1-\beta_l)(L_{i+1}-L_i)<0\\
\dots\\
R_1-\beta_r( V + (1-\beta_1)( R_2-L_1+L_2 ) + \\ (1-\beta_1)(1-\beta_2)( R_3-L_2+L_3 ) + \dots+\\ \prod_{l=i-k}^{i-1}(1-\beta_l)(R_i-L_{i-1}+L_i)+\prod_{l=i-k}^{i}(1-\beta_l)(L_{i+1}-L_i ) <0
\end{cases}
\end{equation}

where k satisfies $1<k<i-1$. The requirements for reputation which makes the victim pay the ransoms for the first i rounds and refuse to pay in the (i+1)-th round are obtained from $b_{i+1}=L_{i+1}, b_i=R_i+\beta_iL_i+(1-\beta_i)b_{i+1}$, $b_{i-1}=R_{i-1}+\beta_{i-1}L_{i-1}+(1-\beta_{i-1})b_{i},\dots ,b_2=R_2+\beta_2L_2+(1-\beta_2)b_{3},b_1=R_1-\beta_r(V + \beta_1L_1 +(1-\beta_1)b_2)+(1-\beta_r)L_1$. For the remaining reputation parameters $\beta_{i+1},\dots,\beta_n$, they have no contribution to the objective function, so it is valid to temporarily ignore these parameters when solving for the optimal expected profit. Furthermore the constraint $b_{i+1}=L_{i+1}$ is easy to be satisfied (by choosing $\beta_{i+1}=1$). Then one can write down the expected profit for the attacker for all the cases $(2\le i<n)$. Let $EP_{a.k}$ denote the expected utility for the attacker when the victim chooses to pay the first $k$ ransoms and refuses to pay the succeeding ransom.

\begin{equation}\label{expected_i_profit}
\begin{split}
    EP_{a,n} = &R_1  + (1-\beta_r)A_1 + \beta_r( -C_r + \beta_1A_1 + \\&(1-\beta_1)( R_2 + \beta_2A_2 + (1-\beta_2)( R_3+\cdots ) ) )\\
    =& R_1 + A_1  - \beta_rC_r +\beta_r\sum_{l=1}^{n-2}(R_{l+1}+A_{l+1}-A_l)\prod_{s=1}^l(1-\beta_s) + \\&\beta_r\prod_{s=1}^{n-1}(1-\beta_{s})(R_n-A_{n-1}) + \beta_r\prod_{s=1}^{n-1}(1-\beta_{s})\beta_nA_n\\
    EP_{a,i} =&R_1 + (1-\beta_r)A_1 + \beta_r( -C_r + \beta_1A_1 + \\&(1-\beta_1)( R_2 + \beta_2A_2 + (1-\beta_2)( R_3+\cdots ) ) )\\
    =&  R_1 + A_1  - C_r  +\beta_r\sum_{l=1}^{n-2}(R_{l+1}+A_{l+1}-A_l)\prod_{s=1}^l(1-\beta_s) + \\&\beta_r\prod_{s=1}^{i-1}(1-\beta_{s})(R_i+A_i-A_{i-1}) + \beta_r(A_{i+1}-A_i)\prod_{s=1}^{i}(1-\beta_{s})\beta_n\\
    EP_{a,1} =&R_1 + A_1 - \beta_r C_r  + (1- \beta_r)(A_2-A_1) \\
\end{split} 
\end{equation}
Using the following substitution, one can transform the aforementioned optimization problem of maximizing the expected utility \ref{expected_i_profit} under the constraints of the inequality system \ref{inequalty_con_ofiransom} into a linear programming problem.
\begin{equation}
    \begin{cases}
        x_1 = \beta_r , \\
        x_2 = \beta_r(1-\beta_1), \\
        x_3 = \beta_r(1-\beta_1)(1-\beta_2), \\
        \qquad \vdots \\
        x_n = \beta_r(1-\beta_1)(1-\beta_2)\dots(1-\beta_{n-1}), \\
        x_{n+1} = \beta_r(1-\beta_1)(1-\beta_2)\dots(1-\beta_{n-1})(1-\beta_n)
    \end{cases}
\end{equation}

In the case where the victim pays the first $i$ ransom, the problem finding the optimal reputation transforms into the following linear programming problem.
\begin{equation}
\begin{aligned}
& \max_{x_1, \dots, x_{i+1}} \quad && -C_rx_1 + \sum_{l=2}^{i} ( R_l - A_{l-1} + A_l ) x_l + x_{i+1}(A_{i+1}-A_i)\\
& \text{s.t.} \quad &&  R_ix_{i} + (L_{i+1}-L_i)x_{i+1} < 0 \\
& && R_{i-1}x_{i-1} +  ( R_i-L_{i-1}+L_i )x_i - L_ix_{i+1} < 0   \\
& && \dots \\
& && R_{i-k}x_{i-k} + \sum_{l=1}^{k}(R_{i-k+l}+L_{i-k+l}-L_{i-k+l-1})x_{i-k+l}   \\
& && - L_ix_{i+1}  <0 \qquad  1<k<i-1 \\
& &&\cdots\\
& && R_1 - Vx_1 -  \sum_{l=1}^{i-1}(R_{i-l}+L_{i-l}-L_{i-l-1})x_{l} + L_ix_{i+1} < 0  \\
& && 0\le x_{n+1}\le x_{n}\le \cdots \le x_{1} \le 1.
\end{aligned}
\end{equation}
The maximum reputation point for the linear programming problem above is the optimal reputation for the attacker when the victim pays the first $i$ ransoms and refuses to pay the succeeding ransom. Then let's consider the remaining cases (the victim pays all the rounds and the victim only pays the first ransom). When the victim pays all the payment rounds, the constraints for the reputation can be written as the follows.
\begin{equation}\label{inequalty_con_ofiransom_always_pay}
\begin{cases}
R_n -(1- \beta_n) L_n < 0  \\
R_{n-1}  + (1-\beta_{n-1})(R_n-L_{n-1}+\beta_nL_n) <0 \\
\cdots\\
R_{i-k} + (1-\beta_{i-k})( R_{i-k+1} - L_{i-k}+L_{i-k+1})+\\(1-\beta_{i-k})(1-\beta_{i-k-1})( R_{i-k+2} - L_{i-k+1}+L_{i-k+2})+\dots+\\ \prod_{l=i-k}^{i-1}(1-\beta_l)(R_i-L_{i-1})+\prod_{l=i-k}^{i-1}(1-\beta_l)\beta_iL_i<0\\
\cdots\\
R_1+L_1-\beta_r( V + (1-\beta_1)( R_2-L_1+L_2 ) + \\ (1-\beta_1)(1-\beta_2)( R_3-L_2+L_3 ) + \dots+\\ \prod_{l=i-k}^{i-1}(1-\beta_l)(R_i-L_{i-1})+\prod_{l=i-k}^{i-1}(1-\beta_l)\beta_iL_i ) <L_1
\end{cases}
\end{equation}

The only difference is the first constraint from $b_n=R_n+\beta_nL_n$. The first constraint has been changed into $R_nx_{n}-L_nx_{n-1}<0$. The other parts including the objective function have almost the same form as (\ref{inequalty_con_ofiransom}) and (\ref{expected_i_profit}). If the victim only pay the first ransom, then the constraint will turn to:
\begin{equation}\label{inequalty_con_ofiransom_only_pay_first}
R_1 - \beta_r V + \beta_1L_1+(1-\beta_1)L_2<L_1
\end{equation}

Except $\beta_1$, the selection of other parameters only needs to ensure that the victim refuses to pay in the remaining n-1 rounds (for instance, by setting $\beta_2$=1). Under this scenario, the other parameters will not affect the expected payoff in (\ref{expected_i_profit}). This way, one can derive the optimal attacker reputation in each region and the corresponding maximum expected payoff. 

The corresponding optimal reputation will be $(x_1^{(l)},1-\frac{x_2^{(l)}}{x_1^{(l)}},\dots,1-\frac{x_{l+1}^{(l)}}{x_l^{(l)}})$ where $I_1$, $I_l$, and $I_n$ are defined below.

\allowdisplaybreaks
\begin{align}\label{op_pro}
I_n=& \max_{x_1, \dots, x_{i+1}} \quad && -C_rx_1 + \sum_{l=2}^{n} ( R_l - A_{l-1} + A_l ) x_l - x_{n+1}A_n\\
& \text{s.t.} \quad &&  R_nx_{n} - L_nx_{n+1} < 0 \\
& && R_{n-1}x_{n-1} +  ( R_n-L_{n-1}+L_n )x_n - L_nx_{n+1} < 0   \\
& && \dots \\
& && R_{n-k}x_{n-k} + \sum_{l=1}^{k}(R_{n-k+l}+L_{n-k+l}-L_{n-k+l-1})x_{n-k+l}   \\
& && - L_nx_{n+1}  <0 \qquad  1 \le k<n-1 \\
& &&\cdots\\
& && R_1 - Vx_1 -  \sum_{l=2}^{n}(R_{l}+L_{l}-L_{l-1})x_{l} + L_nx_{n+1} < 0  \\
& && 0\le x_{n+1}\le x_{n}\le \cdots \le x_{1} \le 1.\\
& &&\cdots \\
I_i=& \max_{x_1, \dots, x_{i+1}} \quad && -C_rx_1 + \sum_{l=2}^{i} ( R_l - A_{l-1} + A_l ) x_l + x_{i+1}(A_{i+1}-A_i)\\
& \text{s.t.} \quad &&  R_ix_{i} + (L_{i+1}-L_i)x_{i+1} < 0 \\
& && R_{i-1}x_{i-1} +  ( R_i-L_{i-1}+L_i )x_i +(L_{i+1}- L_i)x_{i+1} < 0   \\
& && \dots \\
& && R_{i-k}x_{i-k} + \sum_{l=1}^{k}(R_{i-k+l}+L_{i-k+l}-L_{i-k+l-1})x_{i-k+l}   \\
& && -(L_{i+1}- L_i)x_{i+1}  <0 \qquad  1<k<i-1 \\
& &&\cdots\\
& && R_1 - Vx_1 -  \sum_{l=1}^{i-1}(R_{i-l}+L_{i-l}-L_{i-l-1})x_{l}\\
& &&+ (L_{i+1}- L_i)x_{i+1} < 0  \\
& && 0\le x_{i+1}\le x_{n}\le \cdots \le x_{1} \le 1.\\
& &&\cdots \\
I_1=& \max_{x_1, \dots, x_{i+1}} \quad && -C_rx_1 + (A_2-A_1)x_2\\
& \text{s.t.} \quad &&  R_1 - Vx_1 + (L_2-L_1)x_2 < 0 \\
& && 0\le x_{2} \le x_{1} \le 1.\\
\end{align}

Then, by taking the maximum among them, one can obtain the globally optimal attacker reputation and the maximum expected payoff. 
\end{proof}

\subsection{Proof of Theorem~\ref{th6}}

\begin{proof}
   First prove that when the ransom amount for the first round is less than $V$, the victim will definitely choose to pay. According to theorem \ref{th3}, one can obtain the victim's decision in the first round from $b_1$. 
   \begin{equation}
       \begin{split}
           b_1 &= \min\{ R_1 + (1-\beta_r)L_1 + \beta_r(-V +\beta_1L_1+(1-\beta_1)b_2) , L_1 \} \\
           &=\min \{ R_1+L_1+\beta_r( -V + (1-\beta_1)(b_2-L_1)) , L_1 \}\\
           &\le \min\{ (1-\gamma)V + \beta_r( -V+(1-\beta_1)(L_2-L_1)) ) , 0 \} + L_1\\
       \end{split}
   \end{equation}
   If $\beta_r>1-\gamma$, then the victim will choose to pay the first ransom. Its profit is at least $(1-\gamma )V+A_1$ with the second element reputation $\beta_1=1$. In this case, the profit is larger than the profit when the victim refuses to pay. 
   Then one can calculate the lower bound of the expected attacker profit, which can be obtained through solving the final optimization problem defined in \ref{op_pro}. 
   \begin{equation*}
       \begin{aligned}
Pro > R_1+A_1+ I_1 =&  \max_{x_1, \dots, x_{i+1}} \quad && R_1+A_1 -C_rx_1 + (A_2-A_1)x_2\\
& \text{s.t.} \quad &&  R_1 - Vx_1 + (L_2-L_1)x_2 < 0 \\
& && 0\le x_{2} \le x_{1} \le 1.\\
\ge& R_1+A_1-C_r
\end{aligned}
   \end{equation*}
The lower bound is the expected profit when the attacker returns the decryption key but sells the data immediately after.
\end{proof}

%% file: ref.bib
@article{FANG2022102685,
title = {Determination of ransomware payment based on Bayesian game models},
journal = {Computers \& Security},
volume = {116},
pages = {102685},
year = {2022},
issn = {0167-4048},
doi = {https://doi.org/10.1016/j.cose.2022.102685},
url = {https://www.sciencedirect.com/science/article/pii/S0167404822000839},
author = {Rui Fang and Maochao Xu and Peng Zhao}
}

@article{10.1145/3691340,
author = {McIntosh, Timothy and Susnjak, Teo and Liu, Tong and Xu, Dan and Watters, Paul and Liu, Dongwei and Hao, Yaqi and Ng, Alex and Halgamuge, Malka},
title = {Ransomware Reloaded: Re-examining Its Trend, Research and Mitigation in the Era of Data Exfiltration},
year = {2024},
issue_date = {January 2025},
publisher = {Association for Computing Machinery},
address = {New York, NY, USA},
volume = {57},
number = {1},
issn = {0360-0300},
url = {https://doi.org/10.1145/3691340},
doi = {10.1145/3691340},
journal = {ACM Comput. Surv.},
month = oct,
articleno = {18},
numpages = {40}
}

@InProceedings{10.1007/978-3-540-45146-4_8,
author="Camenisch, Jan
and Shoup, Victor",
editor="Boneh, Dan",
title="Practical Verifiable Encryption and Decryption of Discrete Logarithms",
booktitle="Advances in Cryptology - CRYPTO 2003",
year="2003",
publisher="Springer Berlin Heidelberg",
address="Berlin, Heidelberg",
pages="126--144",
isbn="978-3-540-45146-4"
}

@misc{cryptoeprint:2018/187,
      author = {Sean Bowe and Ariel Gabizon},
      title = {Making Groth's zk-{SNARK} Simulation Extractable in the Random Oracle Model},
      howpublished = {Cryptology {ePrint} Archive, Paper 2018/187},
      year = {2018},
      url = {https://eprint.iacr.org/2018/187}
}

@misc{cryptoeprint:2016/260,
      author = {Jens Groth},
      title = {On the Size of Pairing-based Non-interactive Arguments},
      howpublished = {Cryptology {ePrint} Archive, Paper 2016/260},
      year = {2016},
      url = {https://eprint.iacr.org/2016/260}
}

@inproceedings {291054,
author = {Changchang Ding and Yan Huang},
title = {Dubhe: Succinct {Zero-Knowledge} Proofs for Standard {AES} and related Applications},
booktitle = {32nd USENIX Security Symposium (USENIX Security 23)},
year = {2023},
isbn = {978-1-939133-37-3},
address = {Anaheim, CA},
pages = {4373--4390},
url = {https://www.usenix.org/conference/usenixsecurity23/presentation/ding-changchang},
publisher = {USENIX Association},
month = aug
}

@article{9895237,
  author={Alzahrani, Saleh and Xiao, Yang and Sun, Wei},
  journal={IEEE Access}, 
  title={An Analysis of Conti Ransomware Leaked Source Codes}, 
  year={2022},
  volume={10},
  number={},
  pages={100178-100193},
  doi={10.1109/ACCESS.2022.3207757}
}

@misc{fde2023,
  title = {fde},
  author = {PopcornPaws},
  year = {2023},
  url = {https://github.com/PopcornPaws/fde},
  note = {Accessed 2025-05-29}
}

@inproceedings{saver,
author = {Lee, Jiwon and Choi, Jaekyoung and Kim, Jihye and Oh, Hyunok},
title = {SAVER: SNARK-Compatible Verifiable Encryption},
year = {2025},
isbn = {978-3-031-78678-5},
publisher = {Springer-Verlag},
address = {Berlin, Heidelberg},
booktitle = {Financial Cryptography and Data Security: 28th International Conference, FC 2024, Willemstad, Cura\c{c}ao, March 4–8, 2024, Revised Selected Papers, Part II},
pages = {209–226},
numpages = {18}
}

@inproceedings{cao2023hash,
  title={A Hash Time Lock Mechanism Based on Threshold Algorithm},
  author={Cao, Ling and Bao, Pengcheng},
  booktitle={2023 3rd International Conference on Computer Science and Blockchain (CCSB)},
  pages={95--100},
  year={2023},
  organization={IEEE}
}

@article{10.1145/3514229,
author = {Oz, Harun and Aris, Ahmet and Levi, Albert and Uluagac, A. Selcuk},
title = {A Survey on Ransomware: Evolution, Taxonomy, and Defense Solutions},
year = {2022},
issue_date = {January 2022},
publisher = {Association for Computing Machinery},
address = {New York, NY, USA},
volume = {54},
number = {11s},
issn = {0360-0300},
url = {https://doi.org/10.1145/3514229},
doi = {10.1145/3514229},
journal = {ACM Comput. Surv.},
month = sep,
articleno = {238},
numpages = {37}
}

@misc{cryptoeprint:2017/540,
      author = {Jens Groth and Mary Maller},
      title = {Snarky Signatures: \\ Minimal Signatures of Knowledge from Simulation-Extractable {SNARKs}},
      howpublished = {Cryptology {ePrint} Archive, Paper 2017/540},
      year = {2017},
      url = {https://eprint.iacr.org/2017/540}
}

@misc{AES_zero_knowledge_proof_circuit,
  title = {AES zero knowledge proof circuit},
  author = {Lambdaclass},
  year = {2022},
  url = {https://github.com/lambdaclass/AES_zero_knowledge_proof_circuit},
  note = {Accessed 2025-05-29}
}

@article{vacca2021systematic,
  title={A systematic literature review of blockchain and smart contract development: Techniques, tools, and open challenges},
  author={Vacca, Anna and Di Sorbo, Andrea and Visaggio, Corrado A and Canfora, Gerardo},
  journal={Journal of Systems and Software},
  volume={174},
  pages={110891},
  year={2021},
  publisher={Elsevier}
}

@misc{ransomcost,
author= {{Cyber Security Ventures}},
title= {Global Ransomware Damage Costs Predicted To Exceed \$275 Billion By 2031},
url = {https://cybersecurityventures.com/global-ransomware-damage-costs-predicted-to-reach-250-billion-usd-by-2031/},
year={2025}
}

@inproceedings{ransomdecision2pay2018,
author = {Zarifis, Alex and Cheng, Xusen},
year = {2018},
month = {08},
pages = {},
title = {The Impact of Extended Global Ransomware Attacks on Trust: How the Attacker's Competence and Institutional Trust Influence the Decision to Pay},
booktitle = "Proceedings of the Americas
Conference on Information Systems"
}

@InProceedings{10.1007/978-3-319-94782-2_7,
author="Caporusso, Nicholas and Chea, Singhtararaksme and Abukhaled, Raied",
editor="Ahram, Tareq Z.  and Nicholson, Denise",
title="A Game-Theoretical Model of Ransomware",
booktitle="Advances in Human Factors in Cybersecurity",
year="2019",
publisher="Springer International Publishing",
address="Cham",
pages="69--78",
isbn="978-3-319-94782-2"
}

@InProceedings{10.1007/978-3-030-17656-3_12,
author="Fisch, Ben",
editor="Ishai, Yuval
and Rijmen, Vincent",
title="Tight Proofs of Space and Replication",
booktitle="Advances in Cryptology -- EUROCRYPT 2019",
year="2019",
publisher="Springer International Publishing",
address="Cham",
pages="324--348",
isbn="978-3-030-17656-3"
}

@inproceedings{10.1145/1655008.1655015,
author = {Bowers, Kevin D. and Juels, Ari and Oprea, Alina},
title = {Proofs of Retrievability: Theory and Implementation},
year = {2009},
isbn = {9781605587844},
publisher = {Association for Computing Machinery},
address = {New York, NY, USA},
url = {https://doi.org/10.1145/1655008.1655015},
doi = {10.1145/1655008.1655015},
booktitle = {Proceedings of the 2009 ACM Workshop on Cloud Computing Security},
pages = {43–54},
numpages = {12},
location = {Chicago, Illinois, USA},
series = {CCSW '09}
}

@InProceedings{10.1007/978-3-642-17373-8_11,
author="Kate, Aniket
and Zaverucha, Gregory M.
and Goldberg, Ian",
editor="Abe, Masayuki",
title="Constant-Size Commitments to Polynomials and Their Applications",
booktitle="Advances in Cryptology - ASIACRYPT 2010",
year="2010",
publisher="Springer Berlin Heidelberg",
address="Berlin, Heidelberg",
pages="177--194",
isbn="978-3-642-17373-8"
}

@misc{atomicfairdataexchange,
      author = {Ertem Nusret Tas and István András Seres and Yinuo Zhang and Márk Melczer and Mahimna Kelkar and Joseph Bonneau and Valeria Nikolaenko},
      title = {Atomic and Fair Data Exchange via Blockchain},
      howpublished = {Cryptology {ePrint} Archive, Paper 2024/418},
      year = {2024},
      doi = {10.1145/3658644.3690248},
      url = {https://eprint.iacr.org/2024/418}
}

@misc{morgan2019ransomware,
  author       = {Morgan, Steve},
  year         = {2019},
  title        = {Global Ransomware Damage Costs Predicted to Reach \$20 Billion (USD) by 2021},
  url = {https://cybersecurityventures.com/global-ransomware-damage-costs-predicted-to-reach-20-billion-usd-by-2021/},
  note         = {Cybersecurity Ventures, October 21, Accessed June 29, 2018}
}

@misc{kaspersky2016security,
  author       = {Kaspersky},
  year         = {2016},
  title        = {Kaspersky security bulletin 2016},
  url = {https://securelist.com/kaspersky-security-bulletin-2016-story-of-the-year/76757/},
  note         = {Accessed June 29, 2018}
}

@article{li2021game,
  title={Game Theory of Data-selling Ransomware},
  author={Li, Zhen and Liao, Qi},
  journal={Journal of Cyber Security and Mobility},
  pages={65--96},
  year={2021}
}

@misc{6reasonsnottopay,
    author = {Dale Shulmistra},
    year = {2024},
    title = {6 Reasons Not to Pay the Ransom in a Ransomware Attack (Updated with Negotiation Tips)},
    url = {https://invenioit.com/security/pay-the-ransom/} 
}

@article{o2018evolution,
  title={Evolution of Ransomware},
  author={O'Kane, Philip and Sezer, Sakir and Carlin, Domhnall},
  journal={Iet Networks},
  volume={7},
  number={5},
  pages={321--327},
  year={2018},
  publisher={Wiley Online Library}
}

@article{gazet2010comparative,
  title={Comparative Analysis of Various Ransomware Virii},
  author={Gazet, Alexandre},
  journal={Journal in computer virology},
  volume={6},
  pages={77--90},
  year={2010},
  publisher={Springer}
}

@inproceedings{li2020ransomware,
  title={Ransomware 2.0: to Sell, or Not to Sell a Game-theoretical Model of Data-selling Ransomware},
  author={Li, Zhen and Liao, Qi},
  booktitle={Proceedings of the 15th International Conference on Availability, Reliability and Security},
  pages={1--9},
  year={2020}
}

@article{goldreich1994definitions,
  title={Definitions and Properties of Zero-knowledge Proof Systems},
  author={Goldreich, Oded and Oren, Yair},
  journal={Journal of Cryptology},
  volume={7},
  number={1},
  pages={1--32},
  year={1994},
  publisher={Springer}
}

@article{Cartwright2019ToPO,
  title={To Pay or Not: Game Theoretic Models of Ransomware},
  author={Edward J. Cartwright and Julio C{\'e}sar Hern{\'a}ndez Castro and Anna Cartwright},
  journal={J. Cybersecur.},
  year={2019},
  volume={5},
  pages={tyz009},
  url={https://api.semanticscholar.org/CorpusID:157063917}
}

@techreport{Pagnia1999OnTI,
  title={On the Impossibility of Fair Exchange without a Trusted Third Party},
  author={Pagnia, Henning and G{\"a}rtner, Felix C and others},
  year={1999},
  institution={Citeseer}
}

@inproceedings{wang2018overview,
  title={An Overview of Smart contract: Architecture, Applications, and Future trends},
  author={Wang, Shuai and Yuan, Yong and Wang, Xiao and Li, Juanjuan and Qin, Rui and Wang, Fei-Yue},
  booktitle={2018 IEEE Intelligent Vehicles Symposium (IV)},
  pages={108--113},
  year={2018},
  organization={IEEE}
}
